# An Imaging Search for Post-Main-Sequence Planets of Sirius B

Miles Lucas,[1] Michael Bottom,[2] Garreth Ruane,[3] and Sam Ragland[4]

[1]*Institute for Astronomy, University of Hawaii at Manoa, 2680 Woodlawn Dr, Honolulu, HI 96822, USA*
[2]*Institute for Astronomy, University of Hawaii at Hilo, 640 N Aohoku Pl, Hilo, HI 96720, USA*
[3]*Jet Propulsion Laboratory, California Institute of Technology, 4800 Oak Grove Dr, Pasadena, CA 91109, USA*
[4]*W.M. Keck Observatory, 65-1120 Mamalahoa Hwy, Waimea, HI 96743, USA*

## ABSTRACT

We present deep imaging of Sirius B, the closest and brightest white dwarf, to constrain post-main-sequence planetary evolution in the Sirius system. We use Keck/NIRC2 in L′-band (3.776 µm) across three epochs in 2020 using the technique of angular differential imaging. Our observations are speckle-limited out to 1 AU and background-limited beyond. The $5\sigma$ detection limits from our best performing epoch are 17 to 20.4 L′ absolute magnitude. We consider multiple planetary formation pathways in the context of Sirius B's evolution to derive mass sensitivity limits, and achieve sub-Jupiter sensitivities at sub-AU separations, reaching $1.6\,\mathrm{M_J}$ to $2.4\,\mathrm{M_J}$ at 0.5 AU down to a sensitivity of $0.7\,\mathrm{M_J}$ to $1.2\,\mathrm{M_J}$ at >1 AU. Consistent with previous results, we do not detect any companions around Sirius B. Our strong detection limits demonstrate the potential of using high-contrast imaging to characterize nearby white dwarfs.

## 1. INTRODUCTION

In recent decades thousands of exoplanets have been discovered orbiting stars that will eventually leave the stability of the main-sequence (MS) (Akeson et al. 2013). The eventual fate of planets around these stars is uncertain due to the large expansion, stellar winds, and high irradiation encountered during giant branch evolution (Veras 2016). Despite this, evidence from white dwarf pollution (Jura et al. 2007; Xu & Jura 2012), debris disks (de Ruyter et al. 2006; Zuckerman et al. 2010; Koester et al. 2014), and substellar companions (e.g., Luhman et al. 2011; Vanderburg et al. 2020; Blackman et al. 2021) culminate to suggest planetary systems beyond the MS are more common than previously thought. Discovery and characterization of post-MS planets is essential to study how they transform and how they interact with their hosts during the most critical phases of stellar evolution.

The post-MS lifetimes of intermediate-mass stars ($1\,\mathrm{M_\odot}$ to $8\,\mathrm{M_\odot}$) comprises a relatively brief period of giant branch evolution before all nuclear fusion ends and the stars become white dwarfs. As the main-sequence star runs out of hydrogen to burn in its core, it expands to hundreds of times its size, engulfing any companions within the stellar radius. When helium fusion ignites, the giant star becomes three to four orders of magnitude brighter, causing stellar winds and strong irradiation. Eventually, the star runs out of fuel and concludes its nuclear burning, becoming a white dwarf. The white dwarf begins cooling, becoming three to four orders of magnitude dimmer than its MS progenitor.

There is limited knowledge of planetary systems around evolved stars. The pathway for a *first-generation* planet to survive around a post-MS host is violent and uncertain. During giant branch evolution, the planet needs to escape engulfment as well as tidal shredding from its inflated host (Burleigh et al. 2002; Nordhaus & Spiegel 2013). Planets which survive their hosts' inflation are privy to the effects of stellar winds and the high luminosities of asymptotic giant stars (Mustill & Villaver 2012; Mustill et al. 2013; Veras 2016). The stellar winds will chemically enrich the circumstellar environment with metals and dust, and the high luminosity and proximity to the stellar surface will cause strong irradiation, and therefore heating (Spiegel & Madhusudhan 2012). The mass-loss from the stellar winds can adiabatically expand a planet's orbit (Jeans 1924), potentially destabilizing the orbit and chaotically eject-

Corresponding author: Miles Lucas
mdlucas@hawaii.edu

ing the planet (Kratter & Perets 2012). Numerical simulations suggest that a giant planet needs to be $\gtrsim 5\,\mathrm{AU}$ from a solar-like host to escape expansion and tidal effects (Spiegel & Madhusudhan 2012; Nordhaus & Spiegel 2013). When combined with adiabatic orbit expansion, this creates a "forbidden" region of exoplanet phase space for orbital separations closer than $\sim 10\,\mathrm{AU}$.

Recent discoveries of exoplanets in "forbidden" formation regions (Vanderburg et al. 2020; Blackman et al. 2021) suggest evidence for a class of *second-generation* companions. Perets (2010, 2011) describe a planetary formation pathway where, in multi-star systems, the stellar ejecta from an evolving giant star forms a proto-planetary disk around another star (or, in fact, the whole system). These disks serve as reservoirs of material and energy for planet formation, which could be kick-started by a first-generation planet acting as a seed. Such disks would have lifetimes of 1 Myr to 100 Myr which is commensurate with both gravitational-instability ("hot start") and core-accretion ("cold start") formation theories (Marley et al. 2007; Spiegel & Burrows 2012). Another formation pathway considers the chaotic evolution of companion orbits due to stellar mass loss in the presence of multiple bodies. Perets & Kratter (2012) describe this interaction for triplet systems in detail (the "triple evolution dynamical interaction", or TEDI). Kratter & Perets (2012) explore similar dynamical interactions in the restricted three-body problem and concluded up to $\sim 10\%$ of all white dwarf binaries might contain "star-hopper" planets which migrate between the stars.

Previous searches for substellar companions around white dwarfs (e.g., Debes & Sigurdsson 2002; Hogan et al. 2009; Luhman et al. 2011; Xu et al. 2015) have primarily focused on detecting wide-orbit planets which survived the giant branch evolution of their hosts. Recent evidence and theoretical works (e.g., Xu & Jura 2012; Koester et al. 2014; Veras 2016; Vanderburg et al. 2020), though, suggest "forbidden" regions of planetary evolution are worth investigating for exotic post-MS planets. These planets would provide crucial insight into planetary system evolution and planet-star interactions during giant branch evolution.

Direct imaging is well-suited for finding planets around white dwarfs due to the intrinsic faintness of the host (compared to a MS star) and lack of spectral features for radial velocity studies (Burleigh et al. 2002). With direct imaging, exoplanets can be detected and characterized independently from the orbital period of the planet, the stellar radius, or stellar spectrum, which makes it powerful compared to broadband spectral-energy distribution analysis or transit photometry methods. High-contrast imaging pushes direct imaging to its limits using large-aperture telescopes, adaptive optics, coronagraphy, low-noise detectors, image-processing, and observational techniques. Despite the power of high-contrast imaging, very few observations have been carried out for white dwarfs (a recent example is Pathak et al. 2021).

Exoplanets are faint enough that in certain cases only a few photons per second reach the detector. Host stars are typically many orders of magnitude brighter than the thermal emission of giant planets ($\sim 10^{-8}$), and planets have very small angular separations from their hosts, which makes it difficult to disentangle the planet-light from the stellar diffraction pattern (Traub & Oppenheimer 2010). In addition, atmospheric seeing and instrumental aberrations greatly reduce the sensitivity to exoplanets due to the manifestation of quasi-static perturbations of the instrumental point-spread function (speckles, see Guyon 2018). Adaptive optics (AO) corrects a large percentage of the effects of speckles but has decreasing efficacy for dim targets due to the photon-limited nature of modern AO instrumentation. AO is also paramount for enabling coronagraphy, which attenuates on-axis starlight while transmitting off-axis signal, up to some inner working angle.

Typical high-contrast targets are nearby young objects observed in the near-infrared. Younger planets are brighter, thanks to their latent formation heat (Fortney et al. 2010) and observing them in the near-infrared corresponds with their peak blackbody emission. Nearby targets have larger angular separations between potential planets and their hosts, which makes it easier to separate the planet from the star. Nearby targets are also intrinsically brighter, enabling effective AO control. White dwarfs are atypical high-contrast targets due to their age and sparsity, but their relative faintness is compelling for the reduced star-planet flux ratio. In addition, a post-MS planet would be much younger than the system age of a white dwarf and would therefore still retain a large portion of its latent heat, further motivating high-contrast searches of nearby white dwarfs.

In this work we set out to perform high-contrast observations of Sirius B to search for post-MS planets. In the rest of this report, we will introduce the Sirius system as a potential candidate for post-MS planets, along with previous studies of white dwarf Sirius B (Section 2). We will detail our 2020 near-infrared observations of Sirius B with Keck/NIRC2, as well as our processing steps and statistical analysis for companion detection (Sections 3 to 4). Lastly, we will discuss our results within the context of Sirius and post-MS systems, as well as future directions for post-MS imaging (Sections 5 to 6).

## 2. SIRIUS B AND THE SIRIUS SYSTEM

The Sirius system is the 7th closest to the sun at 2.7 pc, consisting of Sirius A, a $J = -1.36$ magnitude A1V star, and Sirius B, a DA white dwarf with a 50 yr orbit (Gaia Collaboration et al. 2016; Bond et al. 2017; Gaia Collaboration et al. 2021). As mentioned previously, the proximity and intrinsic faintness of Sirius B (compared to a MS star) make it compelling for direct imaging. Additionally, the young system age ($\sim$225 Myr) means any giant planets would still retain much of their latent formation heat, increasing their luminosity in the IR (Fortney et al. 2010).

Sirius is one of the oldest studied star systems; the breadth and depth of knowledge about the binary gives exceptional precision for characterizing the circumstellar environment. Most recently, Bond et al. (2017) used Hubble Space Telescope (HST) along with old photographic plates to compile the most precise orbital solution for Sirius to date. Their astrometric uncertainties are over an order of magnitude improved from the visual orbit derived by van den Bos (1960). Following the procedure of Gatewood & Gatewood (1978), Bond et al. (2017) derived dynamical masses of $2.063 \pm 0.023\,\mathrm{M_\odot}$ and $1.018 \pm 0.011\,\mathrm{M_\odot}$ for A and B, respectively. A companion around Sirius B would be affected by the orbit of Sirius A, and this constrained three-body system has been studied numerically (Holman & Wiegert 1999). Bond et al. (2017) calculated the longest stable companion period around Sirius B to be 1.8 yr, which corresponds to a 1.5 AU circular orbit.

The total age of Sirius B is the combination of its white dwarf cooling age and the time from the zero-age main sequence (ZAMS) to the tip of the giant branch (TGB). We adapt the cooling age derived by Bond et al. (2017, Sec. 8) of 126 Myr. We use the updated white dwarf initial-final mass relation (IFMR) of Cummings et al. (2018) to estimate the Sirius B progenitor mass of $5.1 \pm 1.1\,\mathrm{M_\odot}$.

We adopt the system age derived in Cummings et al. (2018) using MIST isochrones of 225 Myr, which implies a ZAMS to TGB age of 99 Myr. These age determinations are limited both by the precision of the stellar parameters as well as the stellar evolution model. The spread of ages derived by Cummings et al. (2018) from different models is $\sim$10 Myr. Determining stellar ages is challenging, and the age uncertainty of $\sim$4% is exceptional compared to those typically obtained by dynamical analysis of young moving groups ($\sim$10%) or gyrochronology ($\sim$15%). The values for our adapted and derived parameters are compiled in Table 1.

One of the peculiarities of the Sirius system is its large eccentricity, $e \sim 0.6$ (Bond et al. 2017). If we assume the orbital expansion due to Sirius B's evolution was adiabatic, we can calculate the initial semi-major axis of the binary

$$a_i = a_f \frac{M_{B,f} + M_{A,f}}{M_{B,i} + M_{A,i}} \quad (1)$$

where $a$ is the system semi-major axis, $M_A$ and $M_B$ are the respective stellar masses, and subscripts $i$ and $f$ correspond to the initial (MS) versus final (post-MS) states (Jeans 1924). The current semi-major axis of the binary is 20 AU, and assuming negligible mass transfer between the two stars, the initial semi-major axis would be $8.6 \pm 1.3$ AU. If the orbit expansion was indeed adiabatic, the eccentricity would be the same before and after evolution. In this case, the periastron of Sirius A and B would be $3.52 \pm 0.52$ AU. Veras (2016) tabulated the maximum stellar radius of intermediate-mass stars during their giant evolution, from which we interpolate a maximum radius for Sirius B of $5.104 \pm 0.075$ AU. This means Sirius A certainly interacted with Sirius B and may have had a common envelope stage. Mass transfer and tidal circularization would be expected; however, the present-day eccentricity provides contrary evidence.

Bonačić Marinović et al. (2008) propose an explanation for the lack of tidal circularization called "tidal-pumping," but neglect to address the observed slow rotation speed of Sirius A (Gray 2014; Takeda 2020), which would be expected to increase with mass transfer to conserve total angular momentum in the binary. Perets & Kratter (2012) suggest the present eccentricity could be due to the chaotic expulsion of a third body between $0.6\,\mathrm{M_\odot}$ to $5.5\,\mathrm{M_\odot}$. Kratter & Perets (2012) point out, though, that the most probable outcome of a planetary-mass companion in a chaotic orbital evolution is a collision with one of the binary components. If Sirius B ejected a first-generation companion during its giant branch evolution, we estimate a $\sim$70% probability of the planet colliding with Sirius A (Kratter & Perets 2012, Fig. 7), although this is far from disqualifying the potential for orbital capture. This is an interesting, although uncertain, explanation for the peculiar surface chemical abundances found on Sirius A (Landstreet 2011; Takeda 2020). The variety in these studies shows the necessity to consider multiple, potentially exotic formation pathways for planetary candidates.

We also consider adiabatic orbit expansion of a substellar companion

$$a_i = a_f \frac{M_f}{M_i} \quad (2)$$

where $a$ is the semi-major axis, and $M$ is the stellar mass of Sirius B. Using the maximum stellar radius of 5.104 ± 0.075 AU and assuming an extra 20% separation to escape tidal shredding (Nordhaus & Spiegel 2013) would create a forbidden region within 31 ± 6 AU around present Sirius B. In combination with the dynamical stability limits of 1.5 AU, we can readily rule out the plausibility of detecting a first-generation companion of Sirius B.

There have been many attempts to find planets in the Sirius system, but, so far, no planets have been detected. Benest & Duvent (1995) suggested the presence of a third body with astrometric perturbations of 100 mas (∼200 $M_J$), but this has so far been unrealized, with Bond et al. (2017) reducing astrometric limits down to 10 mas (∼20 $M_J$). The first modern imaging study searching for companions around Sirius B was Schroeder et al. (2000) who used the HST wide-field planetary camera (WFPC) at 1 µm. Around the same time, Kuchner & Brown (2000) searched in a narrower field of view (FOV) with HST/NICMOS at 1 µm. These studies reported[1] sensitivities down to ∼10 $M_J$ at 5.3 AU (2″). Bonnet-Bidaud & Pantin (2008) used the ground-based ESO/ADONIS instrument in J, H, and Ks-band and reported a sensitivity of ∼30 $M_J$ at 7.9 AU (3″). Skemer & Close (2011) used mid-IR (up to 10 µm) observations from Gemini/T-ReCs, which ruled out evidence for significant infrared excess from massive disks around Sirius B. Thalmann et al. (2011) used Subaru/IRCS at 4.05 µm reporting sensitivities of 6 $M_J$ to 12 $M_J$ at 1″. Recently, Pathak et al. (2021) took coronagraphic mid-IR observations (10 µm) at VLT/VISIR of Sirius A which contained Sirius B in the FOV. Because of the simultaneous observation, their contrast had an azimuthal dependence. Their average reported sensitivity is ∼2.5 $M_J$ at 1 AU, and their best sensitivity (from the "inner" region) is ∼1.5 $M_J$ at 1 AU.

## 3. OBSERVATIONS

We directly imaged Sirius B with Keck/NIRC2 in L′-band (3.776 µm) using the narrow camera (10 mas px$^{-1}$; 2.5″×2.5″) across three epochs in 2020 (Table 2). Despite Sirius B being the brightest white dwarf in the sky, it is still 10 magnitudes fainter than Sirius A, making it a technically challenging target, especially on ground-based telescopes. Our first attempt to observe Sirius B failed due to the light from Sirius A scattering into the FOV of the wavefront sensor (WFS). Vigan et al. (2015, Sec. 2) reported similar issues in their attempts to image Sirius B coronagraphically using VLT/SPHERE. To overcome these obstacles, we decided to use Sirius A as the AO guide star and off-axis guide to Sirius B.

---

[1] A planetary atmosphere and evolution model are needed to derive mass sensitivity limits from imaging. Prior works to our own do not necessarily make the same model choices that we do (Section 5.2), biasing direct comparisons of mass limits.

**Table 1.** Parameters of the Sirius system adopted in this study.

| parameter | value | unit | ref. |
|---|---|---|---|
| $t_{sys}$ | 225 | Myr | B17; C18 |
| $\pi$ | 374.49 ± 0.23 | mas | G21a |
| $d$ | 2.6702 ± 0.0016 | pc | G21a |
| $a$ | 20.016 ± 0.014 | AU | B17 |
| $e$ | 0.591 42 ± 0.000 37 | | B17 |
| Sirius A | | | |
| $M_\star$ | 2.063 ± 0.023 | $M_\odot$ | B17 |
| Sirius B | | | |
| $M_\star$ | 1.018 ± 0.011 | $M_\odot$ | B17 |
| $M_{MS}$ | 5.1 ± 1.1 | $M_\odot$ | B17; C18 |
| $t_{WD}$ | 126 | Myr | B17 |
| $m^{L'}$ | 9.1 ± 0.2 | | BB08 |
| $M^{L'}$ | 11.97 ± 0.20 | | BB08; G21a |

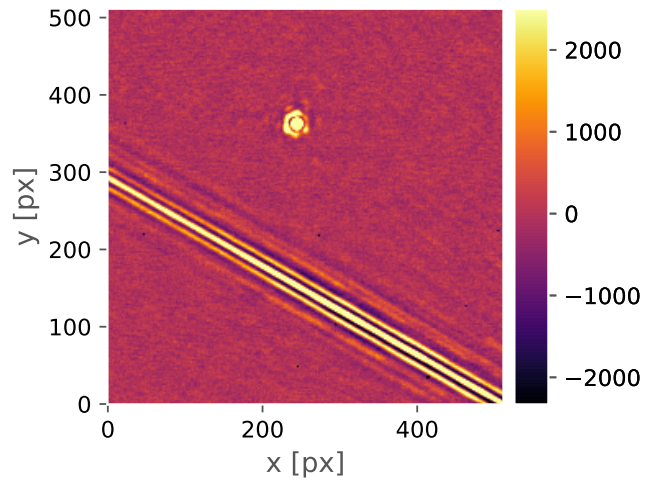

**Figure 1.** Scattered light from Sirius A is present in our FOV around Sirius B as shown by this diffraction spike sweeping across a calibrated science frame of Sirius B from the first epoch. Despite Sirius B's separation of 11″, the overwhelming brightness of Sirius A impedes observations of Sirius B.

**Table 2.** Observing parameters for the three epochs of data. All observations were carried out using the NIRC2 narrow camera (10 mas px$^{-1}$; 2.5″×2.5″) in L′-band (3.776 μm). Observation time is based on the frames that were selected for processing. Seeing values were measured at 0.5 μm using a differential image motion monitor and averaged over the observing session. Seeing values, temperature, and water vapor measurements were all retrieved from the Maunakea weather center forecast archive.

| Date observed | Sirius B offset (″) | Sirius B PA (°) | Obs. time (hr) | FOV rotation (°) | FWHM (mas) | Seeing (″) | Temp (°C) | PWV (mm) |
|---|---|---|---|---|---|---|---|---|
| 2020-02-04 | 11.20 | 67.90 | 1.44 | 60.1 | 79.9 | 0.936 | 0.0 | 0.7 |
| 2020-11-21 | 11.27 | 66.42 | 2.91 | 91.4 | 76.4 | 0.871 | 0.8 | 3.5 |
| 2020-11-28 | 11.27 | 66.38 | 2.44 | 80.4 | 82.2 | 1.23 | -1.5 | 3.0 |

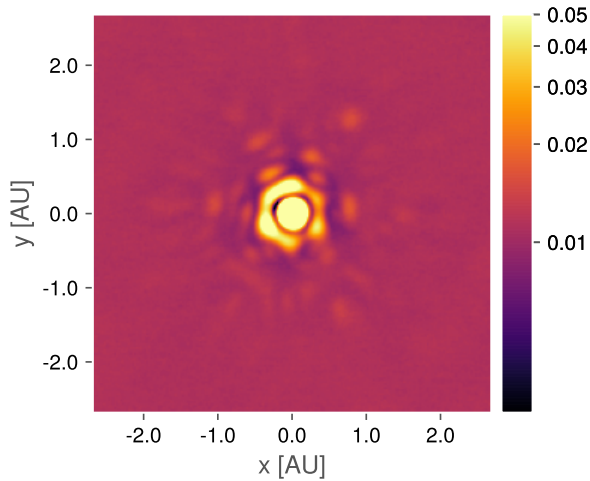

**Figure 2.** The median frame from the second epoch showing the instrumental PSF. The inner core has a Gaussian FWHM of ∼76 mas. The blobs surrounding the first ring are the speckles, with roughly 6-way radial symmetry coinciding with the hexagonal shape of the primary mirror.

Sirius A saturated the WFS of the Keck facility AO system (Wizinowich et al. 2000), so we attenuated the flux using a narrow laser-line filter. While still bright (appearing like a ∼5 magnitude star on the WFS), this was enough attenuation to close the AO loop. From here, we slewed off-axis using the separations and position angles calculated in Table 2 from the orbital solution of Bond et al. (2017). In this mode, we noticed higher than usual drift in the focal plane, requiring manually recentering the target every 5 or 10 minutes. We tried to use the vortex coronagraph (Serabyn et al. 2017) but gave up when the coronagraphic pointing control algorithm, QACITS (Huby et al. 2017), performed erratically in the presence of various diffracted light features. This rendered the coronagraph ineffective, especially with the large amounts of drift. We did not try coronagraphy for the remaining observations.

During each observation, we took dark frames, dome-flat frames, and sky-flat frames for calibration. All observations used the large hexagonal pupil mask and set the telescope's field rotator to track the pupil to exploit the natural rotation of the sky via angular differential imaging (ADI; Marois et al. 2006). To avoid saturation from the sky background, we used 0.4 s integration times and coadded every 75 acquisitions, resulting in ∼30 s per frame in the final images.

## 4. ANALYSIS
### 4.1. Pre-processing

The raw images from NIRC2 required pre-processing before analyzing them for companions. For each epoch, we applied a flat correction using the sky-flat frames captured during observing. We determined the sky-flats had better flat correction than the dome-flats. We also removed bad pixels using a combination of L.A.Cosmic (van Dokkum 2001) and an adaptive sigma-clipping algorithm. We removed the sky background using a high-pass median filter. For both the November epochs we tried exploiting the large focal plane drifts by dithering between two positions to simplify background subtraction, but this ended up performing worse than the high-pass filter. At this point, we manually discarded bad frames, especially those with diffraction spikes from Sirius A within a few hundred pixels, like in Figure 1. Then, we co-registered each frame with sub-pixel accuracy using the algorithm presented in Guizar-Sicairos et al. (2008), followed by fitting each frame with a Gaussian PSF to further increase centroid accuracy.

We centered the co-registered frames in the FOV and cropped them to the inner 200 pixels. With a pixel scale of 10 mas, the crop corresponds to a maximum separation of 1″ or a projected separation of 2.7 AU. We stacked the frames into data cubes for each epoch. We also measured the parallactic angle of each frame, in-cluding corrections for distortion effects following Yelda et al. (2010). For each epoch, we measured the full width at half-maximum (FWHM) of the stellar PSF for use in post-processing by fitting a bivariate Gaussian model to the median frame from each data cube (Figure 2). All the pre-processing code is available in Jupyter notebooks in a GitHub repository (Section 7).

## 4.2. Post-processing

By taking data with the field rotator disabled (ADI), the stellar PSF will not appear to rotate but the FOV will appear to rotate. This allows for the effective separation of companion light from the PSF by spatially decorrelating speckles. After PSF subtraction, we derotated the frames according to their parallactic angle and collapsed the residuals with a variance-weighted sum (Bottom et al. 2017), which reduces the pixel-to-pixel noise as the number of frames in the data cube increases.

For this analysis, we used four ADI algorithms for modeling and subtracting the stellar PSF: median subtraction (Marois et al. 2006), principal component analysis (PCA, also referred to as KLIP; Soummer et al. 2012), non-negative matrix factorization (NMF; Ren et al. 2018), and fixed-point greedy disk subtraction (GreeDS; Pairet et al. 2019b, 2020). We also applied the median subtraction and PCA algorithms in an annular method, where we modeled the PSF in annuli of increasing separation, discarding frames that have not rotated at least 0.5 FWHM (Marois et al. 2006). We used the open-source ADI.jl Julia package for implementations of the above algorithms (Lucas & Bottom 2020).

We determined the best performing PSF subtraction algorithm by measuring the sensitivity to companion signal through repeated injection and recovery of a model PSF. We used a known, fixed S/N for injection to derive the $5\sigma$ detection limits at various positions within the FOV and azimuthally averaged the results to produce a contrast curve. We calculated both the Gaussian contrast and the Student-t corrected contrast, which accounts for the small-sample statistics in each annulus (Mawet et al. 2014). We employed two different detection metrics to search for companions in the residual data: the Gaussian significance map (Mawet et al. 2014) and the standardized trajectory intensity mean map (STIM map; Pairet et al. 2019a). These maps assign the likelihood of a companion to each pixel using different assumptions of the residual statistics. We used ADI.jl for calculating these metrics. The collapsed residual frames along with the above metrics for each algorithm and epoch are in Appendix A.

A common problem when using subspace-driven post-processing algorithms like PCA, NMF, or GreeDS is choosing the size of the subspace (i.e., the number of components). For PCA, NMF, and GreeDS algorithms, we created sets of residual cubes, varying the number of components from 1 to 10. We chose 10 for the maximum number of components because we saw a dramatic decline in contrast sensitivity after the first few components (Figure 12). In our analysis, we employed the STIM largest intensity mask map (SLIM map; Pairet 2020) as an ensemble statistic. The SLIM map calculates the average STIM map from many residual cubes along with the average mask of the $N$ most intense pixels in each STIM map. We expect a real companion to be present in many different residual cubes from the same dataset, so this ensemble statistic gives us a probability map without determining the number of components *a priori*. The collapsed residual frames, average STIM map, SLIM map, and contrast curves for each epoch for each of the above algorithms are in Appendix A. All the code and data used for this analysis is available in a GitHub repository (Section 7).

## 5. RESULTS

We determined the best-performing algorithms for each epoch using the contrast curves described in Section 4. For the first two epochs, full-frame median subtraction had the best contrast at almost all separations. For the last epoch, annular PCA subtraction with 2 principal components and a rotation threshold of 0.5 FWHM produced the best contrast at close separations ($0.2''$ to $0.4''$) and had similar performance to other algorithms beyond $0.4''$. When processed with this algorithm, we were unable to retrieve a 100 S/N injected companion with $5\sigma$ significance in the innermost annulus. The contrast in this region is ill-determined and therefore not plotted. Figures 3 to 4 show the collapsed residual frames from each epoch, along with the Gaussian significance maps and STIM maps.

We show the contrast curves from the best-performing algorithm for each reduction in Figure 5. We determine the limiting sensitivities in terms of the planetary mass by first calculating the contrast-limited absolute magnitude using an L'-band magnitude for Sirius B of 9.1 (adapted from Ks-band magnitude from Bonnet-Bidaud & Pantin 2008) and a distance modulus of $-2.87$ (Gaia Collaboration et al. 2021). We divide Figure 5 into two regimes: speckle-limited and background-limited. The speckle-limited regime exists from $0.2\,\mathrm{AU}$ to $1\,\mathrm{AU}$ characterized by the increasing sensitivity with separation. Here we reach a median $5\sigma$ detection limit of $\sim19$ magnitude (L'). This regime is mainly constrained by

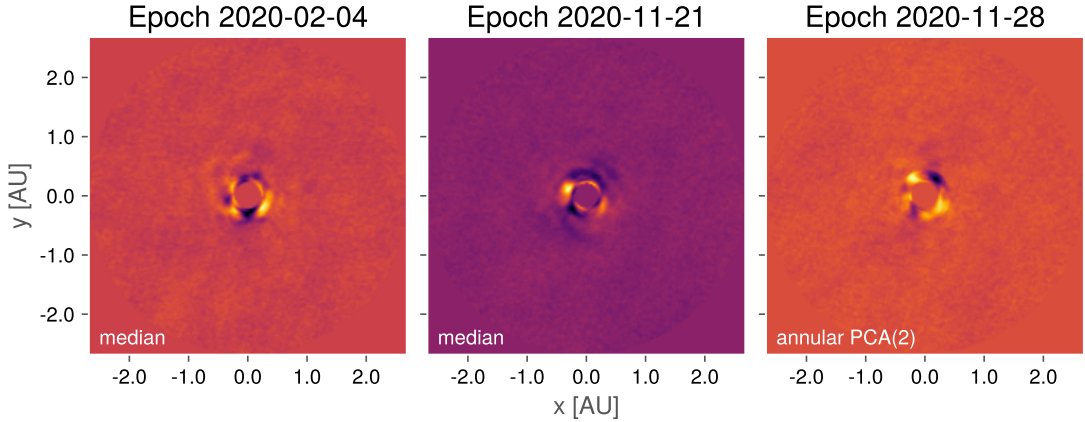

**Figure 3.** The flat residuals of each epoch after PSF subtraction, derotating, and collapsing. The inner two FWHMs are masked out for each frame.

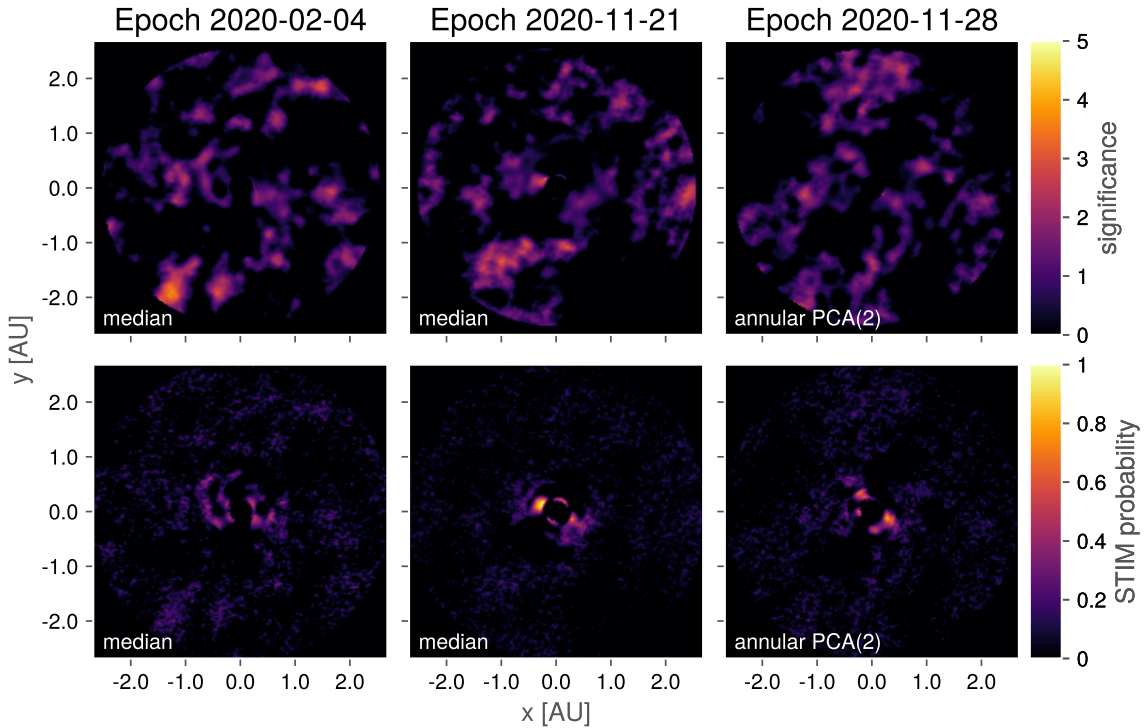

**Figure 4. top row:** The *significance* maps for each epoch accounting for small-sample statistics (Mawet et al. 2014). Typically, a critical value for detection is 5. **bottom row:** The STIM maps for each epoch calculated from each residual cube. The STIM probability has a typical cutoff threshold of 0.5 for significant detections. The inner two FWHMs are masked out for each map.

the quality of the AO correction and the PSF subtraction method. The background-limited regime ($>1\,{\rm AU}$) is characterized by the flattening of the contrast curves and is primarily limited by the sky brightness. In this region, we reach 20.4 magnitude (L′) in the 2020-11-21 epoch. Our data is background-limited due to the relative brightness of the sky in L′(2.91 mag/sq arcsec[2])

[2] https://www2.keck.hawaii.edu/inst/nirc2/sensitivity.html

compared to the pixel-to-pixel noise sources (e.g., read noise).

### 5.1. *Companions around Sirius B*

The reduced images do not show consistent or significant evidence for a substellar companion. The STIM probability maps for the 2020-11-21 and 2020-11-28 epochs suggest evidence for some blobs ∼0.3 AU (0.13″; 1.6 FWHM) from the center. The February epoch also shows a blob at a similar separation in the reduced im-age which does not appear in the STIM map. The lack of statistical evidence in the February epoch and the significance maps as well as the proximity to the central star both reduce the probability of these blobs being true companions. Nonetheless, we estimated astrometry for blobs from each epoch (Table 3) and tried fitting Keplerian orbits using the "Orbits for the Impatient" algorithm (OFTI; Blunt et al. 2017) using the open-source `orbitize` Python package (Blunt et al. 2020). We generated $10^4$ orbits, none of which constrained the points from each epoch (Appendix B). This implies non-Keplerian motion and we take this as direct evidence against the blobs being substellar companions of any kind. We considered the possibility that the blobs are scattered light from a circumstellar debris disk, but this is highly unlikely given the brightness of the blob and the lack of IR excess that such a massive disk would radiate (Skemer & Close 2011). The signal can simply be explained as residual starlight not removed during PSF subtraction.

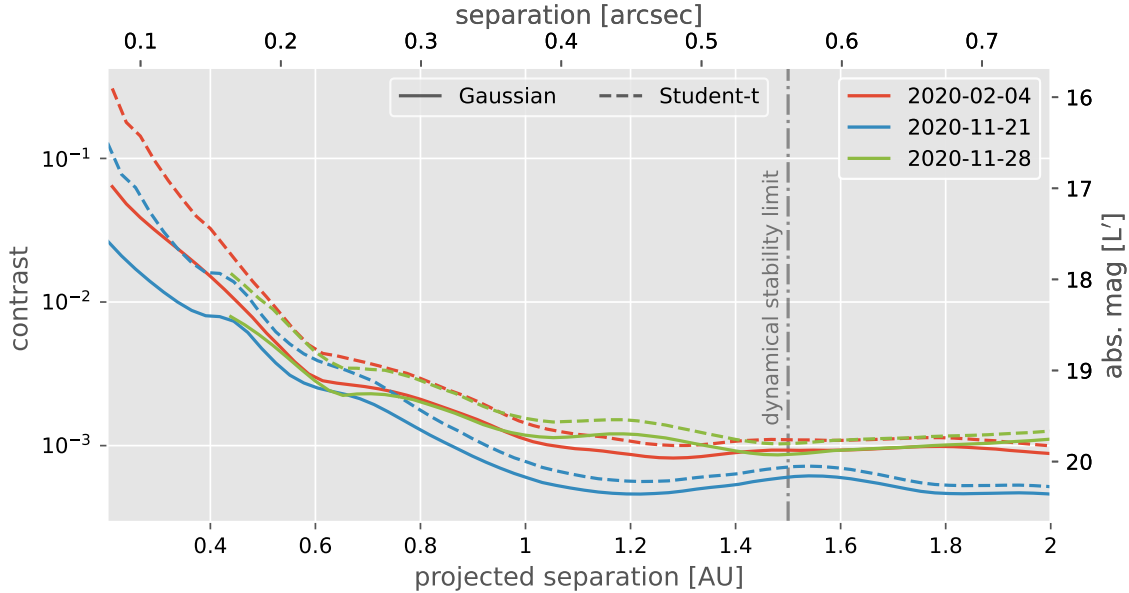

**Figure 5.** The contrast curves for the best performing algorithm from each epoch. The solid lines are the Gaussian $5\sigma$ contrast curves, and the dashed lines are the Student-t corrected curves. The absolute magnitude is calculated using an absolute magnitude for Sirius B of 11.97. The expected upper limit for a dynamically stable orbit of 1.5 AU is plotted as a vertical dashed line. The annular PCA curve cuts off because the innermost annulus was not able to detect a 100 S/N companion with $5\sigma$ significance.

### 5.2. Mass detection limits

To convert our photometric detection limits to mass limits we must employ an appropriate planetary atmosphere model and evolution grid. This is not a trivial task, as the effects of post-MS stellar evolution on planets are highly uncertain and not readily modeled in the currently available grids. In particular, we would like to study the effects of metal and dust enrichment of the circumstellar environment from stellar winds. We used the ATMO2020 model grid (Phillips et al. 2020) with non-equilibrium chemistry due to weak vertical mixing for our solar metallicity model, following Pathak et al. (2021). ATMO2020 models very cool objects better than previous grids such as AMES-Cond (Allard et al. 2012) but are not available for non-solar metallicities. To explore metal enrichment, we employed the Sonora Bobcat grid (Marley et al. 2021a,b) at both solar and +0.5 dex metallicities.

To determine the correct isochrone for the grids, we consider two formation scenarios. If a first-generation planet survived the giant phase of Sirius B through star hopping, it would have an age close to the system age of 225 Myr. If the planet formed in a disk of stellar ejecta during the giant branch evolution, the age would be closer to the white dwarf cooling age of 126 Myr. If the planet formed in such a disk, or if it accreted some of the material, it would almost certainly have peculiar chemistry, although it is uncertain exactly how the relative abundances would change.

Figure 6 shows our most sensitive contrast curve converted to mass limits under the different models. The first panel uses the ATMO2020 models to show how the choice of isochrone age leads to a $\sim 0.3\,M_J$ difference in the background-limited regime. The second panel uses the Sonora models to demonstrate the relatively minor effects ($\sim 0.1\,M_J$) the increased metallicity has on the mass limits. We could not fully utilize the Sonora grid because there are no atmospheric models simulated for effective temperatures below 200 K, which are precisely the models needed for the background-limited regimes. Overall, we constrain our detection limits to $1.6\,M_J$ to $2.4\,M_J$ at 0.5 AU (0.19″) in the speckle-limited regime and ultimately $0.7\,M_J$ to $1.1\,M_J$ at >1 AU (0.38″) in the background-limited regime.

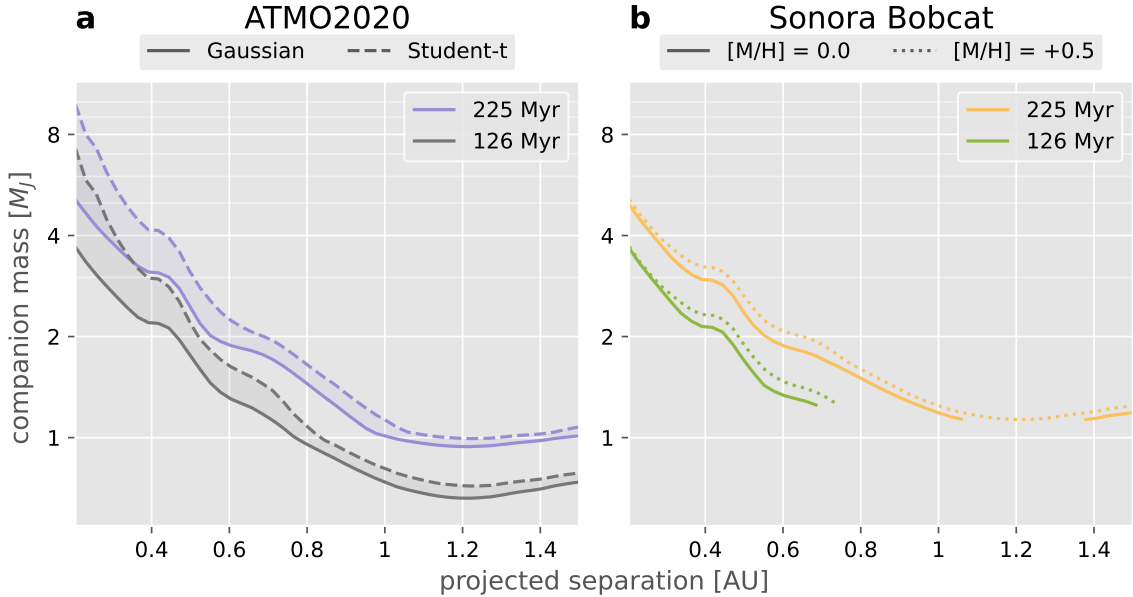

**Figure 6.** Mass sensitivity curves derived from the 2020-11-21 epoch, which has the most sensitive contrast. The limits are calculated from the absolute magnitude derived in the contrast curves. Both curves are truncated at 1.5 AU due to the dynamical stability limit. **(a)** The absolute magnitudes are converted to masses using the ATMO2020 isochrone grid with non-equilibrium chemistry and weak convective mixing. The solid lines are the Gaussian $5\sigma$ detection limits, and the dashed lines are the Student-t corrected limits. The two ages represent the ages of two potential formation pathways, one of which is the system age (225 Myr), the other is the white dwarf cooling age of Sirius B (126 Myr). The relative difference between the ages (first-generation vs. second-generation) comes out to $\sim 0.3\,M_J$ at 1 AU. **(b)** The absolute magnitudes are converted to masses using the Sonora Bobcat grid with solar metallicity (solid lines) and with +0.5 dex metallicity (dotted lines). For clarity, we only show the Gaussian contrast curves in this panel. The Sonora grid does not have atmospheric spectra for $T_{\rm eff} < 200\,K$, which causes the cutoffs around $1\,M_J$. The relative difference due to the metallicity is $\sim 0.1\,M_J$.

## 6. DISCUSSION & CONCLUSIONS

Post-MS planetary evolution has historically been limited to theoretical work. Recently, though, increasingly strong and diverse observational constraints, including new detections, have invigorated the field. We set out in this work to search nearby white dwarf Sirius B for post-MS planets. The Sirius system is one of the most well studied in history, and its precise characterization improves our systematic uncertainties. It is highly unlikely a first-generation planet survived Sirius B's giant branch evolution, but no previous imaging efforts have directly addressed post-MS formation in their analyses.

In this work, we presented high-contrast images of Sirius B in the near-IR. Our $5\sigma$ sensitivity limits are the best that have been reached for Sirius B so far, down to 20.4 L′absolute magnitude at >1 AU. We consider multiple planetary formation pathways yielding ages between 126 Myr to 225 Myr and explore the effects of enriched metallicity. We translate our sensitivity limits using the ATMO2020 and Sonora Bobcat grids to constrain our mass detection limits to $0.7\,M_J$ to $1.1\,M_J$ at >1 AU. Our observations also show how the high precision of the parameters of the Sirius system directly benefits the sensitivity to planets. For example, the $\sim 4\%$ relative age uncertainty (Section 2) translates to a mass uncertainty below $0.1\,M_J$. Despite the high sensitivity of this study, we found no significant evidence for a companion around Sirius B, consistent with previous results.

Although our observations yield no detections, we illustrate the capability of modern high-contrast instrumentation, even without coronagraphy, to reach strong detection limits. Our detection limits benefit directly from the precise stellar characterization of the Sirius system, as well as the proximity and brightness of Sirius B. With laser guide stars (LGS; e.g., van Dam et al. 2006; Baranec et al. 2018) the limiting magnitude for sufficient AO performance is improved ($m^R \lesssim 19$). Future extremely large telescopes will benefit from increased collecting areas and smaller inner working angles due to the larger aperture diameters. The Thirty Meter Telescope (TMT), for example, will have LGS AO in its first-generation instrument suite (NFIRAOS), which will significantly improve the faintness limits of its high-contrast instrumentation. We suspect future work using LGS AO on current and next-generation telescopes will be capable of studying nearby white dwarf systems at the sub-AU and sub-Jupiter-mass scales (Holberg et al. 2016). Such observations could significantly improve our theories of planetary formation and stellar evolution beyond the main sequence.

Future space-based observations with the James Webb Space Telescope (JWST) will avoid the effects of atmospheric seeing and the bright sky background. For example, using JWST/NIRCAM in long-wavelength imaging mode has a limiting magnitude of ∼25 in the F480M filter. The pixel scale ($0.06''\,\text{px}^{-1}$) and PSF size ($\sim 0.3''$) are adequate for sub-AU observations of nearby white dwarfs, depending on the contrast-limited performance of NIRCAM. Unfortunately, Sirius B is far too bright and too close to Sirius A to observe with JWST without severe saturation.

## 7. DATA AND CODE AVAILABILITY

All the code used for pre-processing and reducing the data is available under an open-source license in a GitHub repository.[3] This code includes all of the scripts for generating each figure in this manuscript. The preprocessed data cubes and parallactic angles are available on Zenodo under an open-source license.[4] Inquiries regarding data and code are welcome.

## ACKNOWLEDGMENTS

We thank the anonymous referee for their helpful comments. We thank Michael Liu and Mark Phillips for their expertise and advice on the ATMO2020 model grid. We thank Mark Marley and Didier Saumon for their assistance with the Sonora Bobcat model grid. The data presented herein were obtained at the W. M. Keck Observatory, which is operated as a scientific partnership among the California Institute of Technology, the University of California, and the National Aeronautics and Space Administration. The Observatory was made possible by the generous financial support of the W. M. Keck Foundation. The authors wish to recognize and acknowledge the very significant cultural role and reverence that the summit of Maunakea has always had within the indigenous Hawaiian community. We are most fortunate to have the opportunity to conduct observations from this mountain.

*Facility:* Keck:II (NIRC2)

*Software:* ADI.jl (Lucas & Bottom 2020), astropy (Collaboration et al. 2013; Astropy Collaboration et al. 2018), Julia (Bezanson et al. 2017), numpy (Harris et al. 2020), orbitize (Blunt et al. 2020), scikit-image (Walt et al. 2014),

---

[3] https://github.com/mileslucas/sirius-b
[4] 10.5281/zenodo.5115225

## APPENDIX

### A. ADI PROCESSING RESULTS

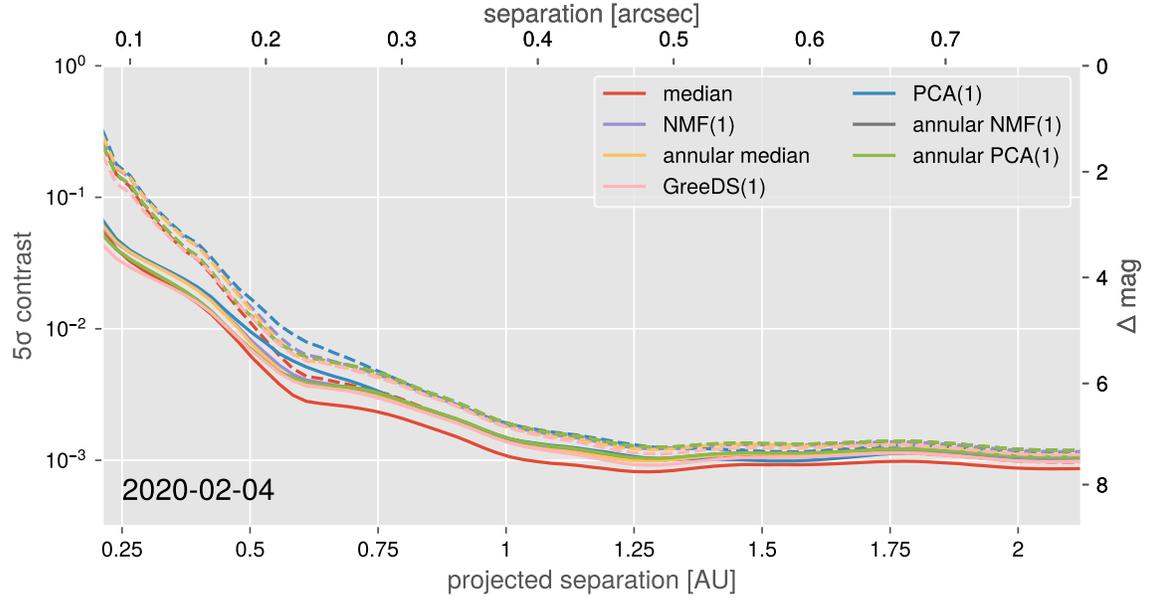

**Figure 7.** $5\sigma$ contrast curves from every ADI algorithm for the first epoch. Both the Gaussian (solid lines) and Student-t corrected (dashed lines) contrast curves are shown.

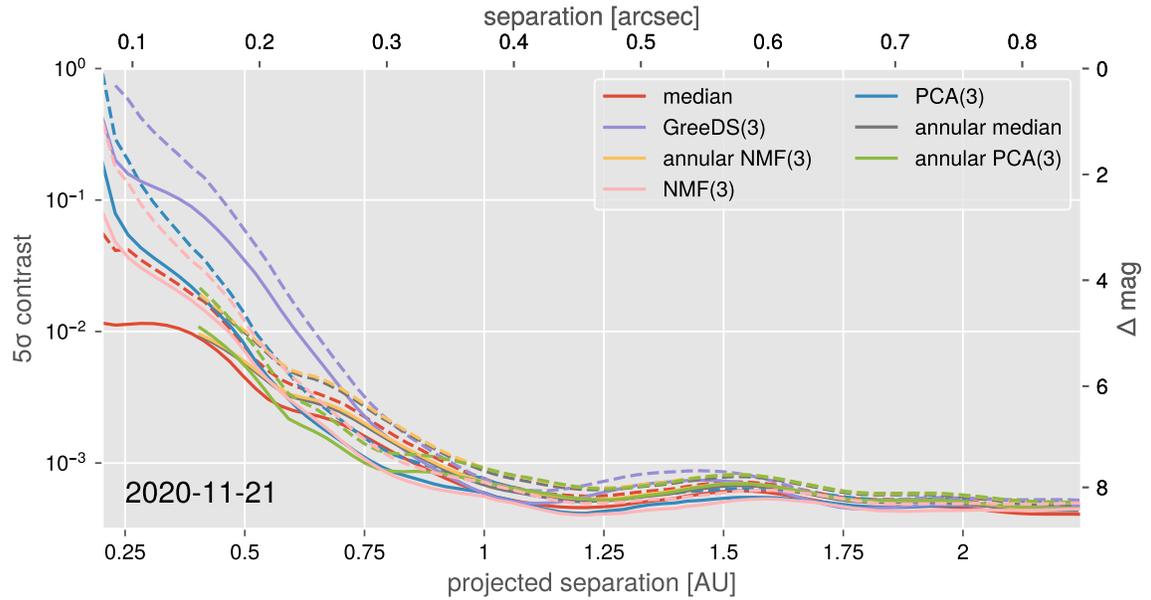

**Figure 8.** $5\sigma$ contrast curves from every ADI algorithm for the second epoch. Both the Gaussian (solid lines) and Student-t corrected (dashed lines) contrast curves are shown.

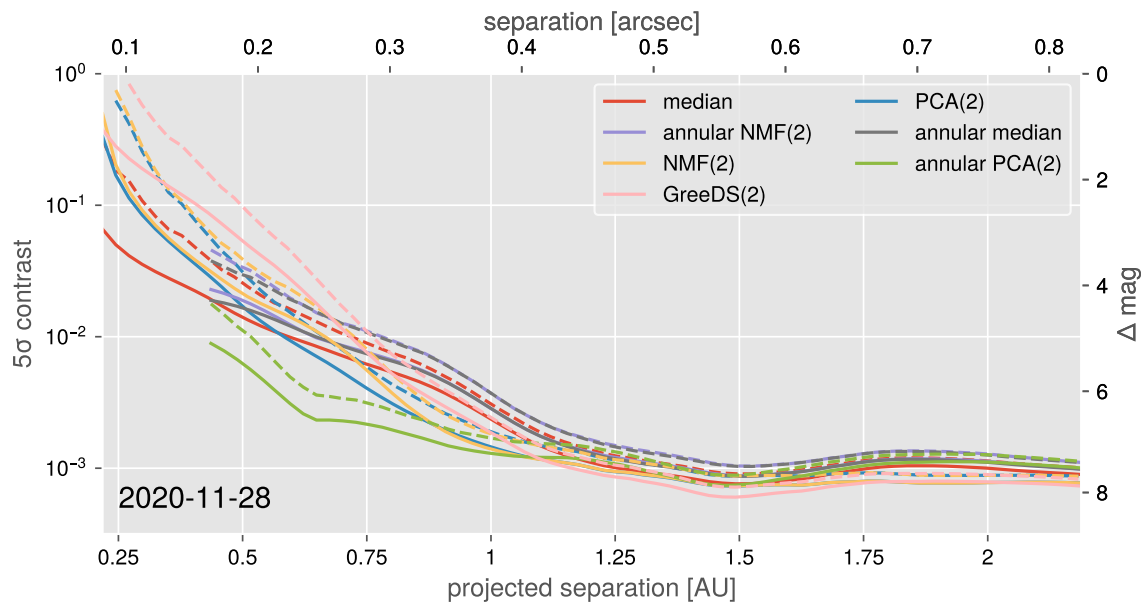

**Figure 9.** $5\sigma$ contrast curves from every ADI algorithm for the third epoch. Both the Gaussian (solid lines) and Student-t corrected (dashed lines) contrast curves are shown.

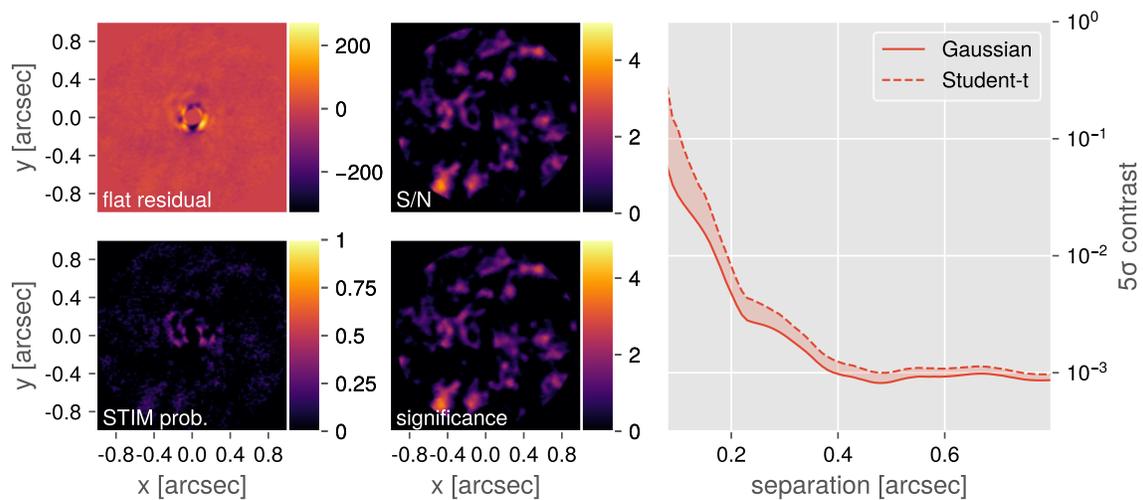

**Figure 10.** Post-processing results from the second epoch using full-frame median subtraction. The top-left frame is the collapsed residual frame, the top-right is the Gaussian S/N map, the bottom-left is the STIM probability map, and the bottom-right is the Student-t corrected significance map. In each frame, the inner two FWHMs are masked out. The right figure show the Gaussian (solid line) and Student-t corrected (dashed curve) $5\sigma$ contrast curve. Outputs for other epochs and other algorithms (21 figures) are in the online figure set and the GitHub repository.

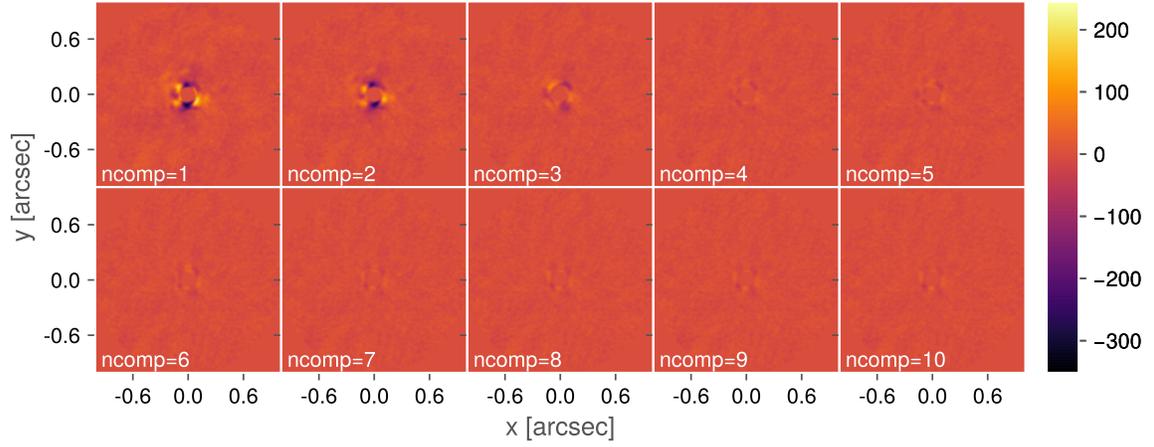

**Figure 11.** Collapsed residual frames from the first epoch using PCA reduction with 1-10 components. The figures share a common scale and the inner two FWHMs are masked out for all the frames. Outputs for the other epochs and the NMF and GreeDS algorithms (9 figures) are in the online figure set and the GitHub repository

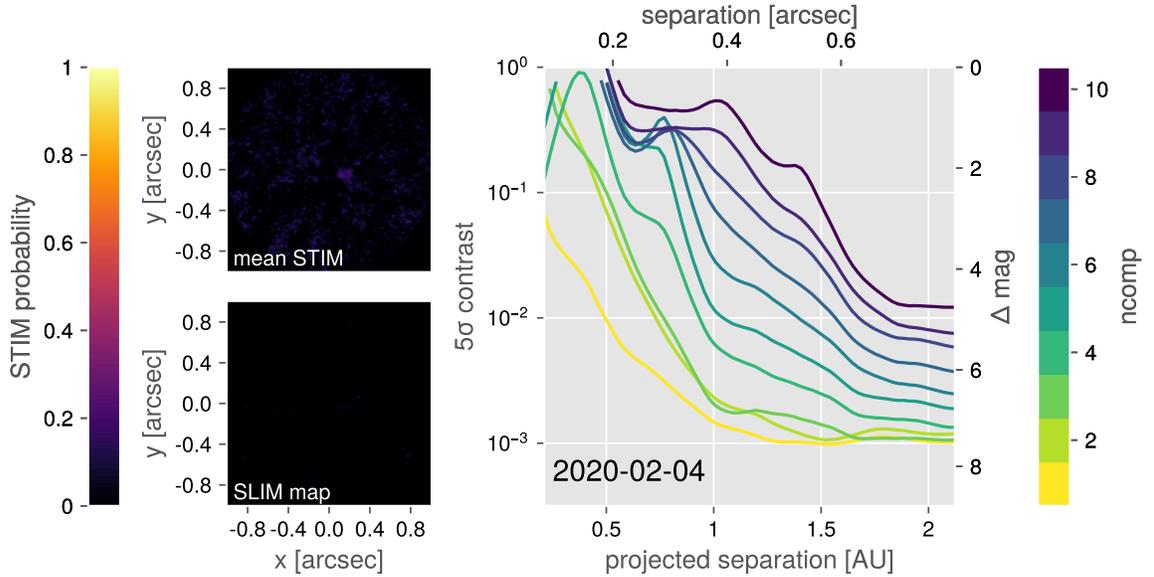

**Figure 12.** $5\sigma$ Gaussian contrast curves for the first epoch using PCA reduction with 1-10 components. The left two figures are the STIM probability map and the SLIM detection map. For both of these maps, a typical cutoff value is 0.5. Outputs for the other epochs and the NMF and GreeDS algorithms (9 figures) are in the online figure set and the GitHub repository.

## B. PROVISIONAL ORBIT FITTING

We found multiple interesting blobs in the reduced data that were not statistically significant. Nonetheless, we tried fitting Keplerian orbits using OFTI (Blunt et al. 2017) to determine the feasibility of the blobs being astrophysical companions. We began by estimating the astrometry of the blobs by eye in the reduced data (Table 3, Figure 13). We tried simulating $10^4$ orbits via rejection sampling with OFTI, but none of the generated orbits contained all three points in their solution. Overall we determined these blobs are not companions and are most likely systematic noise from the stellar PSF.

Table 3. Provisional astrometry for a blob of interest from each epoch. The separation and offset are relative to Sirius B. The uncertainties were derived from the FWHM of the PSF from each epoch.

| Date observed | offset (mas) | PA (°) |
|---|---|---|
| 2020-02-04 | $123 \pm 40$ | $-128 \pm 20$ |
| 2020-11-21 | $119 \pm 38$ | $68 \pm 18$ |
| 2020-11-28 | $132 \pm 41$ | $37 \pm 21$ |

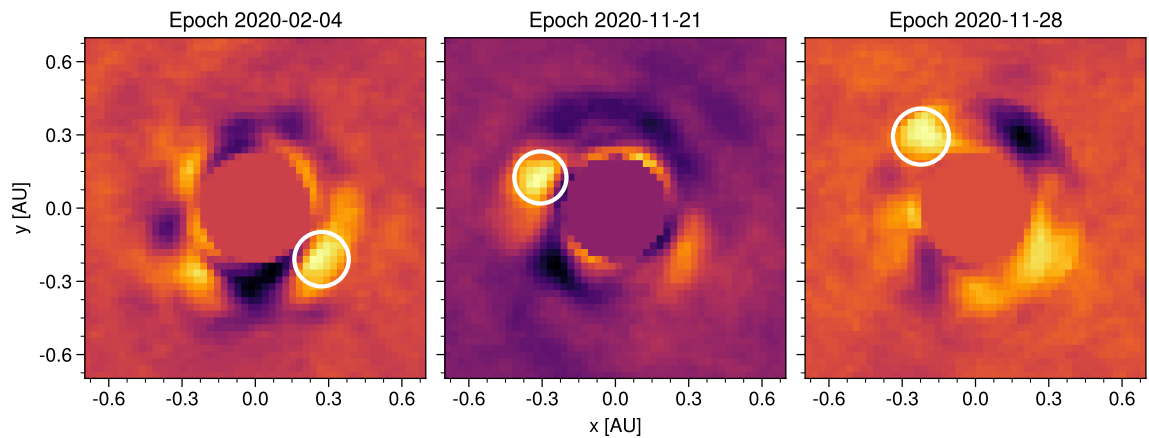

Figure 13. Provisional astrometry (white circles) displayed on collapsed and derotated residual frames from each epoch. Each frame was cropped to the inner $\sim 0.7$ AU ($0.25''$) and the inner two FWHMs have been masked out. The width of the circles represents the measurement uncertainty.